\documentclass{emulateapj}

\begin{document}
\title{Evolution of Galaxy Stellar Mass Functions, Mass Densities, and
  Mass to Light ratios from $z\sim7$ to $z\sim4$}

\author{
Valentino Gonz\'alez\altaffilmark{1},
Ivo Labb\'e\altaffilmark{2,3},
Rychard J. Bouwens\altaffilmark{1},
Garth Illingworth\altaffilmark{1},
Marijn Franx\altaffilmark{4},
Mariska Kriek\altaffilmark{5}
}

\altaffiltext{1}{Astronomy Department, University of California,
Santa Cruz, CA 95064}
\altaffiltext{2}{Carnegie Observatories, 813 Santa Barbara Street,
Pasadena, CA 91101}
\altaffiltext{3}{Hubble Fellow}
\altaffiltext{4}{Leiden Observatory, Leiden University , NL-2300 RA 
Leiden, Netherlands}
\altaffiltext{5}{Department of Astrophysical Sciences, Princeton 
University, Princeton, NJ 08544}

\begin{abstract}

We derive stellar masses from SED fitting to rest-frame optical and UV
fluxes for 401 star-forming galaxies at $z\sim4,~5,$ and 6 from
Hubble-WFC3/IR observations of the ERS combined with the deep GOODS-S
Spitzer/IRAC data (and include a previously-published $z\sim7$
sample).  A mass-luminosity relation with strongly
luminosity-dependent $\mathcal{M}$/L$_{UV}$ ratios is found for the
largest sample (299 galaxies) at $z\sim4$. The relation
$\mathcal{M}\propto\rm{L}_{UV,1500}^{1.7(\pm0.2)}$ has a
well-determined intrinsic sample variance of $0.5\,$dex. This relation
is also consistent with the more limited samples at $z\sim5-7$. This
$z\sim4$ mass-luminosity relation, and the well-established faint UV
luminosity functions at $z\sim4-7$, are used to derive galaxy mass
functions (MF) to masses $\mathcal{M}\sim10^{8}$ at $z\sim4-7$.  A
bootstap approach is used to derive the MFs to account for the large
scatter in the $\mathcal{M}$--L$_{UV}$ relation and the luminosity
function uncertainties, along with an analytical crosscheck.  The MFs
are also corrected for the effects of incompleteness. The
incompleteness-corrected MFs are steeper than previously found, with
slopes $\alpha_M\sim-1.4$ to $-1.6$ at low masses.  These slopes are,
however, still substantially flatter than the MFs obtained from recent
hydrodynamical simulations. We use these MFs to estimate the stellar
mass density (SMD) of the universe to a fixed M$_{\rm{UV,AB}}<-18$ as
a function of redshift and find a SMD growth
$\propto(1+z)^{-3.4\pm0.8}$ from $z\sim7$ to $z\sim4$. We also derive
the SMD from the completeness-corrected MFs to a mass limit
$\mathcal{M}\sim10^{8}$M$_\odot$. Such completeness-corrected MFs and the
derived SMDs will be particularly important for model comparisons as
future MFs reach to lower masses.

\end{abstract}


\section{Introduction} 

Measurements of the stellar mass of high-redshift galaxies
($z\gtrsim4$) provide important constraints on scenarios of galaxy
formation and early evolution. For example, the evidence for a strong
correlation between the observed star formation rate (SFR) and stellar
mass \citep{star09, labb10, labb10a}, with little apparent evolution
between $z\sim3$ and $z\sim7$ \citep{star09, gonz10}, suggests an
epoch of exponential growth which, interestingly, is similar to that
found in simulations \citep[e.g.][]{finl10}.

Recent deep near-IR WFC3/IR observations over the ERS field
\citep{wind10} combined with pre-existing deep GOODS IRAC data provide
access to the rest-frame UV and optical wavelengths of $4<z<7$ star
forming galaxies and hence reasonably accurate estimates of their
$\mathcal{M}$/L ratios and stellar masses. The substantial samples of
$z>4$ galaxies detected with WFC3/IR span a range in stellar mass,
allowing the derivation of mass functions (MFs).

MFs are fundamental characteristics of the galaxy population but in
practice, they are difficult to compute directly, especially at high
redshift because of selection effects, incompleteness, and
contamination by interlopers. An alternative approach to derive the MF
is to start with the well-determined UV LF at these redshifts and
convert to stellar mass using an average $\mathcal{M}$/L
\citep[e.g.][]{mclu09}.  The main advantage is that LFs are corrected
for all selection effects in the data and reach very faint limits,
although such a simple approach does not take into account any
possible luminosity dependence or scatter of the $\mathcal{M}$/L. In
this Letter, we estimate improved MFs at $4<z<7$ from the published
UV-LFs by deriving a relation between UV luminosity and stellar
masses. We also estimate the scatter in the $\mathcal{M}$/L, which
allows us to correct for incompleteness at low stellar masses. We use
the MFs to compute the stellar mass density (SMD) of the universe at
$z=4,~5,~6$ and 7.

We adopt an $\Omega_{M}=0.3$, $\Omega_{\Lambda}=0.7$ cosmology with
$H_0=70~\rm{kms~s^{-1}~Mpc^{-1}}$ throughout.  All magnitudes are in
the AB system \citep{oke83}.

\section{Galaxy Sample from HST and Spitzer Data}

\begin{figure*}
  \centering
  
  \includegraphics[width=1.\textwidth]{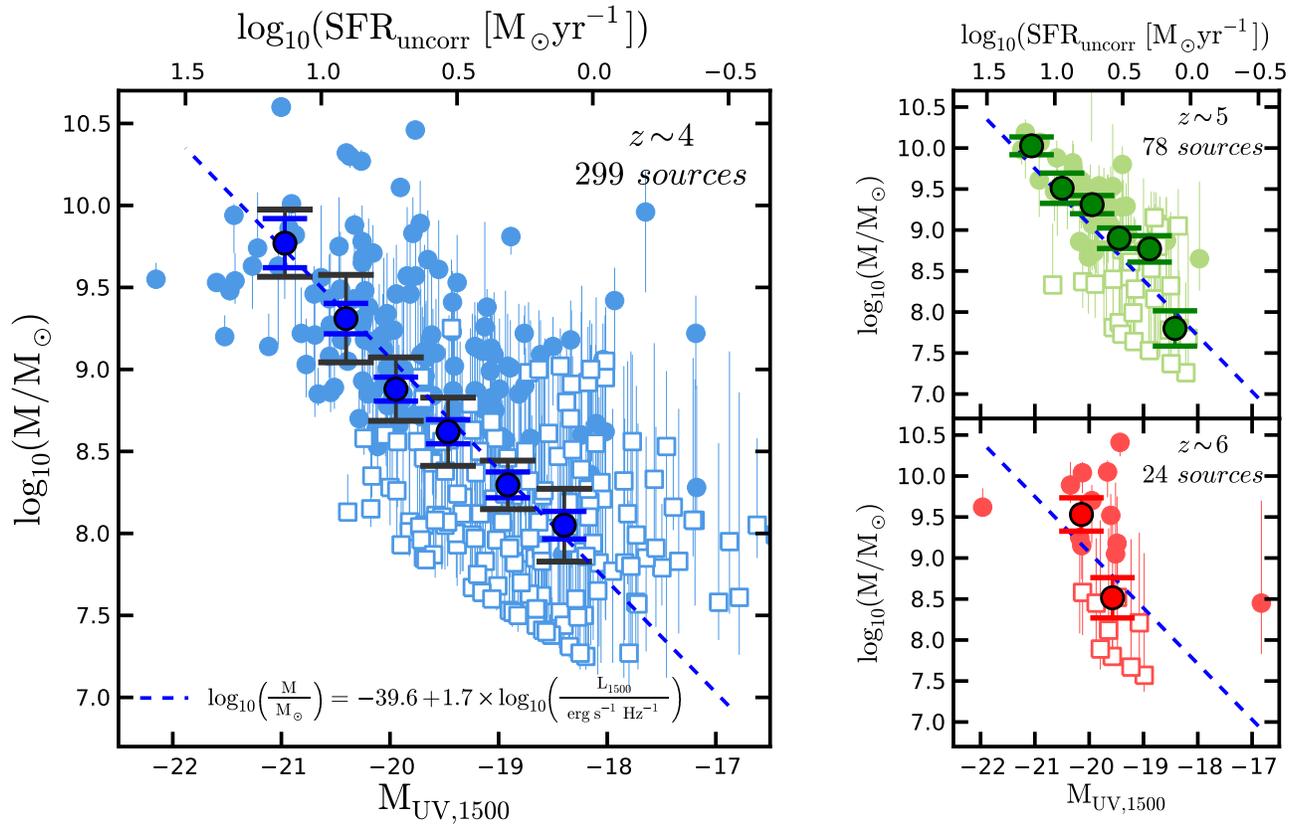}
  
  \caption{Stellar masses as a function of UV Luminosity
    ($\rm{M_{UV,1500}=51.63-2.5\times\log_{10}(L_{UV,1500}[erg~s^{-1}~Hz^{-1}]})$)
    for the $z\sim4~,5,$ and 6 samples. SFR$_{\rm{uncorr}}$ (top axis)
    is derived using the \cite{mada98} conversion formula (no
    extinction correction). The final sample of 401 sources with FAST
    SED-fit mass estimates is shown here.  Open squares indicate
    low-S/N measurements ($<2\,\sigma$ in [3.6]). The darker symbols
    in each panel represent the median mass of the sample
    ($\sim0.5~\rm{M}_{UV,1500}$-mag bins). The small error bars
    represent the bootstrapped errors.  The larger black error bars
    include a conservative estimate of the systematics computed by
    comparing the estimated median mass at a given luminosity with the
    mass estimated from the stacked SEDs at the same luminosity. The
    dashed blue line (slope$~=1.7\pm0.2$) represents the mean trend
    between $\rm{log_{10}\mathcal{M}}$ and $\rm{M_{UV,1500}}$ at
    $z\sim4$. It is consistent with no evolution with redshift. The
    scatter at the luminous end ($\pm0.5\,$dex), where photometric
    errors are small, is intrinsic (see Figure~\ref{mtolvsMUV}).}

\label{mvsMUV}
\end{figure*}

The sources used here for determinations of the $\mathcal{M}$/L ratios
and $z=4-6$ MFs were found in the recent Hubble-WFC3/IR observations
of the ERS field. Both the GOODS ACS optical
($B_{435}V_{606}i_{775}z_{850}$) and the WFC3/IR
($Y_{098}J_{110}H_{160}$) data reach depths of $\sim28$ mag
($5\sigma$, 0.35\arcsec-diameter apertures, see \citealt{bouw10d,
  giav04}).  All sources have Spitzer/IRAC coverage with depths of
27.8 and 27.1 in the [3.6] and [4.5] channels, respectively
($1\,\sigma$ in 2.4\arcsec apertures). The $z\sim7$ sample is taken
from \citep{labb10}.

The $z\sim4,~5,~6$ sample totals 679 objects, consisting of
$524\,B,123\,V,\,$and 32$\,i$-dropouts that were selected with the
same criteria as \cite{bouw07}:
  
  $z\sim4~B$-dropouts:
  $$(B_{435}-V_{606}>1.1)~\wedge~[B_{435}-V_{606}>(V_{606}-z_{850})+1.1]$$
  $$\wedge~(V_{606}-z_{850}<1.6)$$

  $z\sim5~V$-dropouts:
  $$\{[V_{606}-i_{775}>0.9(i_{775}-z_{850})]~\vee~(V_{606}-i_{775}>2)\}$$
  $$\wedge~(V_{606}-i_{775}>1.2)~\wedge~(i_{775}-z_{850}<1.3)$$
  
  $z\sim6~i$-dropouts\footnote{Slightly modified, Bouwens et
    al. 2010, in prep.}:
  $$(i_{775}-z_{850}>1.3)~\wedge~(z_{850}-J_{125}<0.8)$$

The rest-frame optical photometry from Spitzer/IRAC is ideally suited
for deriving stellar masses at these redshifts
\citep[e.g.][]{papo01,yan05,labb10}. A challenge is that the broad
IRAC PSF usually results in these faint sources being contaminated by
foreground neighbors. To obtain reliable IRAC fluxes we use the
deblending method of \citeauthor{labb06} \citep[2006, see
  also][]{gonz10, labb10a, labb10, wuyt07, de-s07}.  Briefly, this
method uses the higher-resolution HST images to create models of both
the foreground neighbors and the source itself. We convolve each model
image with a kernel to simulate the IRAC observations. We fit for all
the sources simultaneously (with independent normalization factors)
and subtract the best fits for the neighbors.  In the clean image of
each dropout we are able to perform standard aperture photometry. We
use 2.5\arcsec-diameter apertures and correct the fluxes to total
assuming stellar profiles ($1.8\times$ in both channels).

As expected, our cleaning procedure does not work for every
source. We restrict our sample to the 60\% of sources with the best
$\chi^2$ residuals. This reduces the number of non-optimal
subtractions to $<8\%$.  The final sample suitable for deriving masses
from the HST+Spitzer data totals 401 sources:
$299~z\sim4,~78~z\sim5$, and 24 at $z\sim6$.  We do not expect this
selection step to introduce any important biases, since it depends on
the distribution of the non-associated neighbors of the source.  Of
the remaining sources, $\sim50\%$ have low IRAC S/N ($<2\,\sigma$ in
[3.6]).

\begin{figure*}
  \centering
  
  \includegraphics[width=1.\textwidth]{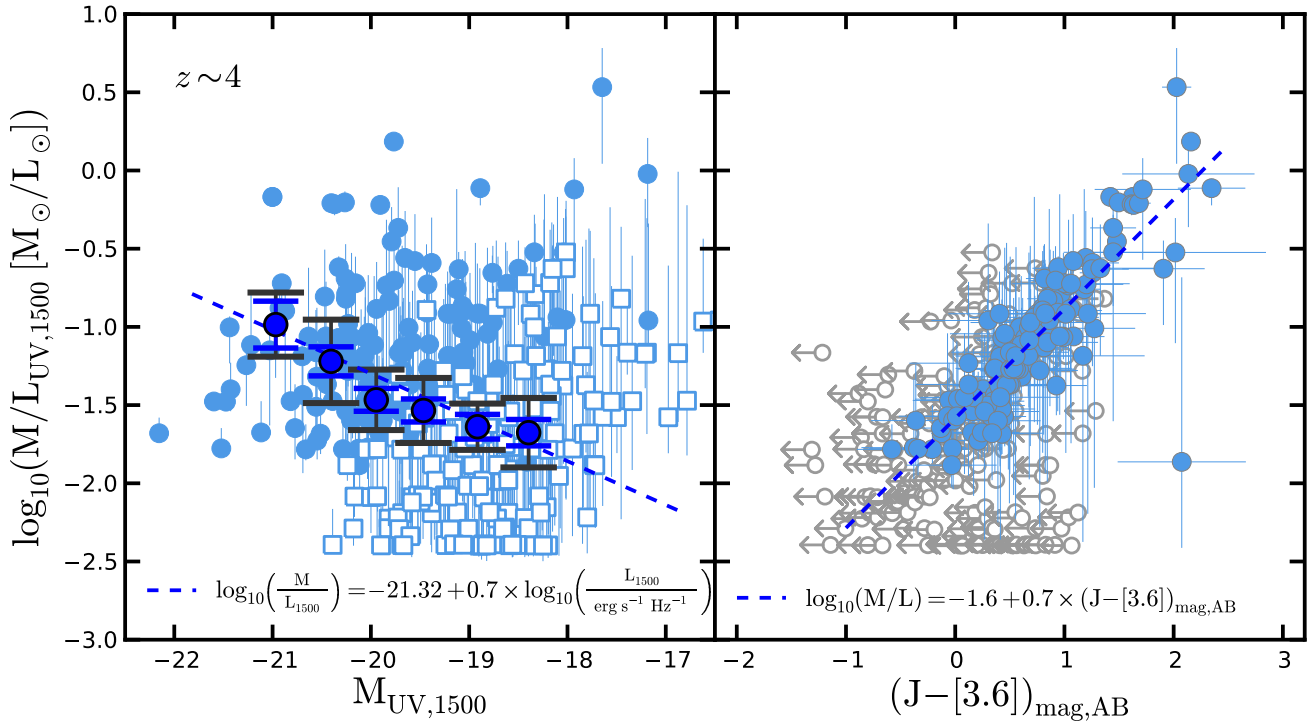}
  
  \caption{(\emph{left}) $\mathcal{M}$/L ratio as a function of UV
    Luminosity for the $z\sim4$ sample. Symbols and error bars as in
    Figure~\ref{mvsMUV}. The median $\mathcal{M}$/L ratio changes by a
    factor $5\times$ in the luminosity range our
    sample. (\emph{right}) The correlation between the $\mathcal{M}$/L
    and $J-[3.6]$ color.  Arrows indicate $2\,\sigma$-upper
    limits. This tight relation suggests that the large scatter
    observed in the $\mathcal{M}$/L (left panel) is largely due to
    intrinsic variations in the UV-to-optical colors.  Photometric
    scatter can only account for $\lesssim0.14\,$dex at
    $\rm{M_{UV,1500}\sim-20}$ ($0.37\,$dex at $-19$). }

  \label{mtolvsMUV}
\end{figure*}

\section{Stellar Mass Estimates from SED Fits}

We use the FAST SED-fitting code \citep{krie09} to derive stellar
masses of the $401~z\sim4-6$ sources from SED fits with the full suite
of fitted parameters. For all sources we fit the broadband
ACS+WFC3/IR+IRAC [3.6] and [4.5] fluxes using the \citet[BC03]{bruz03}
models with a \citet{salp55} initial mass function (IMF, 0.1--100
M$_\odot$) and assuming a 0.2$\,$Z$_\odot$ metallicity.  We also include
the sample of $z\sim7$ galaxies with similarly-determined masses from
\citet{labb10}.

The SFH cannot be uniquely determined from broadband SEDs due to
well-known degeneracies between the star formation timescale, age, and
dust extinction.  We have assumed a SFH with a constant SFR.
Different SFHs introduce systematic offsets to the mass
determinations, largely independent of redshift \citep[cf.][]{papo10}.
The systematic differences between masses based on declining,
constant, or rising SFHs are typically $\lesssim0.3\,$dex
\citep{finl07}.

Figure \ref{mvsMUV} (left) shows the FAST SED-fit stellar masses (from
HST+Spitzer data) versus UV luminosity (bottom axis).  While the
scatter is large (RMS$\sim0.5\,$dex), there is a clear trend of
increasing mass with increasing UV luminosity.  The stellar mass
$\mathcal{M}$--L$_{UV,1500}$ relation at $z\sim4$ is well-fit by
$\rm{log_{10}(\mathcal{M})\propto1.7(\pm0.2)\rm{log_{10}(L_{1500})}}$.
The lower bound that appears at masses $<10^8~\rm{M}_\odot$
corresponds to the $\mathcal{M}$/L of the youngest model we allow
(10$\,$Myrs). This is a reasonable assumption for the majority of the
sample, as IRAC detections in the stacks suggest that continuum is
dominant, not emission lines.  This is also not critical, since our
inferred $\mathcal{M}$/L trend is insensitive to the cutoff.  The
$z\sim4$ relation is consistent with the $z\sim5$ sample, and, in
zero-point, with the small $z\sim6$ sample, and also with the $z\sim7$
sample presented in \citep{labb10}.

Figure \ref{mtolvsMUV} explores the $z\sim4~\mathcal{M}/\rm{L}$ ratio
trend in more detail, showing that
$\mathcal{M}/\rm{L}_{\mathit{UV},1500}$ depends on luminosity; the
$\mathcal{M}/\rm{L}$ ratio is $\sim5\times$ lower at M$_{UV,1500}=-18$
than at M$_{UV,1500}=-21$.  This suggests that UV-faint galaxies
contribute less to the global SMD than assumed in previous studies
\citep{labb10a, labb10}.  However, due to the steep faint-end slope of
the UV LF \citep[e.g.][]{bouw10d}, their high number density makes
their total contribution quite significant.

A striking aspect of the relation is the large scatter in
$\mathcal{M}$/L.  The observed sample variance (one standard
deviation) for our sample is $\sim0.5\,$dex for
$\rm{-21<M_{UV,1500}<-18}$. At the bright end M$_{UV,1500}<-20$ the
scatter is largely intrinsic, whereas at the faint end
M$_{UV,1500}>-19.5$ it is dominated by observational uncertainties. In
particular, the stellar masses of sources with IRAC detections are
much better constrained than IRAC-undetected sources.  Photometric
uncertainties contribute $\sim0.14$ dex to the scatter at
M$_{UV,1500}\sim-20$ (0.37 dex at $-19$).  Moreover, we find that the
$\mathcal{M}$/L ratio is tightly correlated with the $J-[3.6]$ color
(standard deviation 0.18$\,$dex, Figure \ref{mtolvsMUV}, right),
suggesting that the variation is real, and not an artifact of the
modeling.

The relation in Figure \ref{mtolvsMUV} (right) also allows us to
estimate the possible effect of contamination by emission lines (not
included in our models). At $z\sim4$, a $20\%$ contribution of
H$\alpha$ to [3.6] would result in redder $J-[3.6]$ colors and hence
overestimates of the $\mathcal{M}$/L and of the masses by 30\%.  This
would affect the SMDs at all redshifts because they all rely on our
$z\sim4~\mathcal{M}$/L ratio estimates (see \S4).

\section{Stellar Mass Functions at $z\sim4,~5,~6,~\rm{and}~7$.}

\begin{figure*}
  \centering

  \includegraphics[width=.95\textwidth]{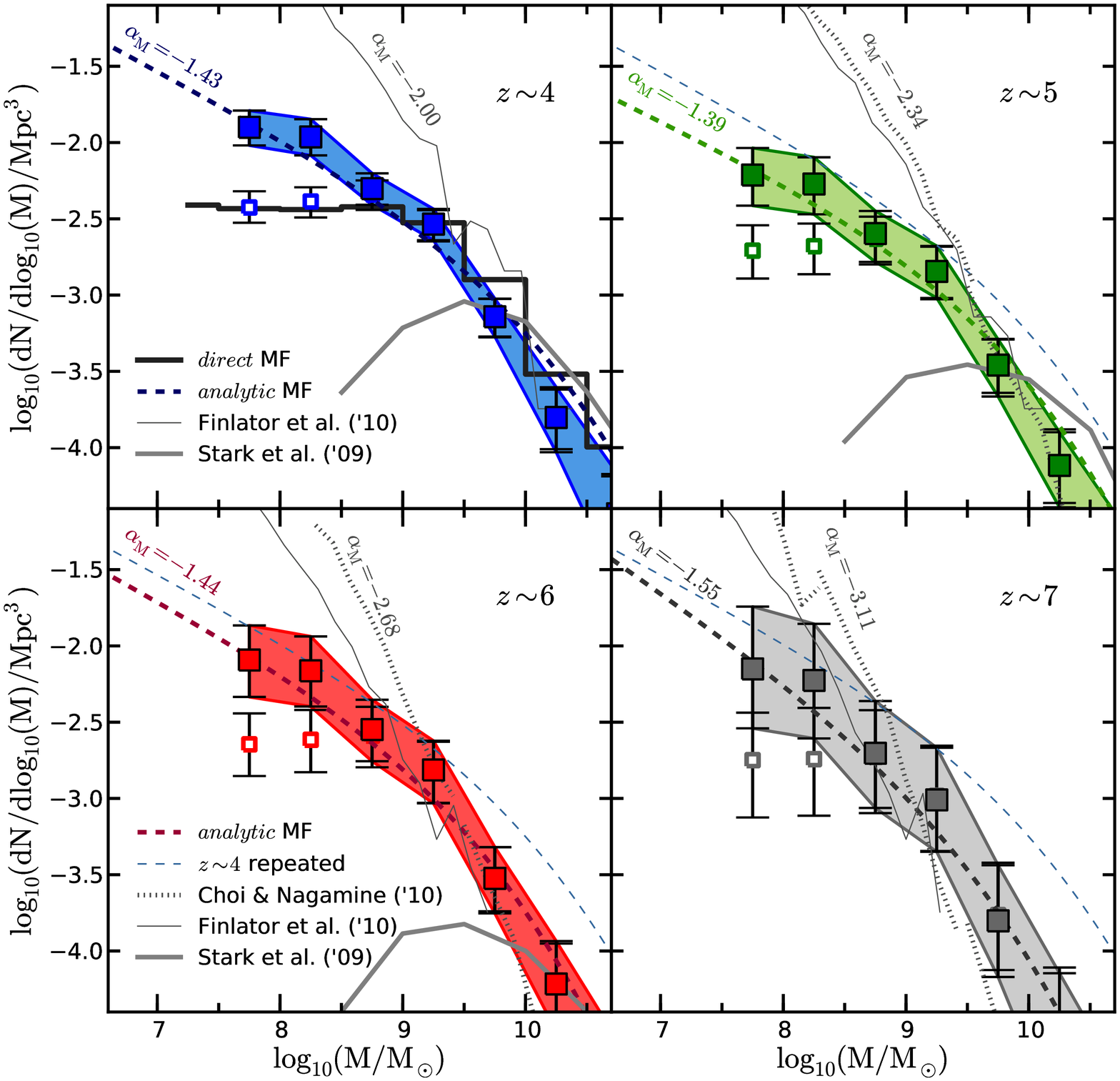}

  \caption{Stellar mass functions at $z\sim4,~5,~6,$ and 7 derived
    from the $\rm{log}(\mathcal{M})-\rm{M_{UV,1500}}$ distribution for
    the $z\sim4~B$-dropouts (Figure$\,$\ref{mvsMUV}), and the
    \cite{bouw07, bouw10d} UV-LFs at $z\sim4-7$. The points are
    derived from the ``bootstrap'' approach (see text).  Errors
    reflect uncertainties in the LF and the $\sim0.5\,$dex $1\,\sigma$
    scatter of the $\mathcal{M}$--M$_{UV,1500}$ relation
    (Figure$\,$\ref{mvsMUV}).  Completeness-corrected values are
    estimated assuming that the $\mathcal{M}$--M$_{UV,1500}$ relation
    extends to fainter limits with similar scatter about the
    extrapolated mean trend (M$_{UV,1500}<-18$ uncorrected: open;
    corrected: filled; dark band is at $1\,\sigma$ around the
    corrected values). The \emph{direct} MF at $z\sim4$ (thick
    histogram) is in good agreement with the uncorrected MF (see
    text).  For masses $>10^{9.5}~\rm{M}_\odot$, the uncorrected $z<7$
    MFs are in rough agreement with the determinations of
    \citet{star09} and of \citet{mclu09} at $z\sim6$ and
    $\mathcal{M}>10^{10}~\rm{M}_\odot$.  The thick dashed curve in
    each panel represents the \emph{analytic} MFs derived from an
    idealized $\mathcal{M}-\rm{{M}_{UV,1500}}$ relation (see text
    \S4). These MFs have low-mass slopes $\alpha_M\sim-1.4--1.6$,
    slightly flatter than the UV LFs
    ($\alpha=-1.7--2.0$:~\citealt{bouw10d}). In turn, the assumed
    symmetric scatter of 0.5$\,$dex flattens their slopes at the
    high-mass end. The $z\sim4$ \emph{analytical} MF is repeated in
    the other panels for comparison (thin dashed curve). The dotted
    and thin solid lines show the simulated MFs from \cite{choi10} and
    Finlator et al. (in preparation). Our new results are corrected
    for incompleteness, yet the difference between our results and the
    simulations is already substantial by
    $\mathcal{M}=10^9~\rm{M}_\odot$. The source of the disagreement is
    unclear.}

  \label{MFfig}
\end{figure*}

The stellar MFs at a given redshift can be estimated by combining the
LF at that redshift with an appropriate $\mathcal{M}$/L
relation. Since UV LFs have been derived from large samples to very
faint limits and carefully corrected for a wide range of potential
biases, they constitute an excellent basis for determining MFs. The
calibration of log$(\mathcal{M})$ vs M$_{UV,1500}$ in Figures
\ref{mvsMUV} and \ref{mtolvsMUV} provides the means to transform the
$z\sim4-7$ UV-LFs into MFs.

However, the scatter about the mean $\mathcal{M}$--M$_{UV,1500}$
relation is so large that ignoring it would produce significant
errors.  Galaxies with relatively low luminosity but high
$\mathcal{M}$/L ratios, for example, contribute significantly at the
high mass end of the MFs. Hence, we take care to determine the average
$\mathcal{M}$--M$_{UV,1500}$ relation in a robust way, we characterize
the scatter at the high mass end and use this estimate of the scatter
at lower luminosities/masses where the observational uncertainties
dominate.

We use two approaches to create the MFs.  First, we use the individual
points in Figure \ref{mvsMUV} as representative of the
$\mathcal{M}$--M$_{UV,1500}$ distribution by bootstrap resampling
them.  To correct for incompleteness we add faint sources to the
distribution. This important step increases the low mass slope of the
MFs substantially. Second, we use the fitted
$\mathcal{M}$--M$_{UV,1500}$ relation and an idealized model of its
scatter to produce what we label as ``analytic'' versions of the
MFs. We compare to other estimates as a cross-check.

\emph{Bootstrapped MFs}: We start with the $z\sim4-7,$ UV-LFs of
\citet{bouw07, bouw10d} and draw 40000 luminosities from each LF in
the range $-21.5<\rm{M}_{UV,1500}<-18$. We convert the luminosities to
stellar masses by bootstrap re-sampling from the distribution of
points at $z\sim4$ in Figure \ref{mvsMUV}.  We use the
$z\sim4~\rm{log}(\mathcal{M})-\rm{M}_{\mathit{UV},1500}$ distribution
at all redshifts because it is well-defined over a wide range of
luminosities and is consistent with the relations at other redshifts,
including the $z\sim7$ relation presented in \cite{labb10}.  To
account for the uncertainties in the LFs we perturb their Schechter
parametrizations within the uncertainties and repeat 5000 times. This
``bootstrap'' process results in the uncorrected MFs (Figure
\ref{MFfig} open squares and error bars).

As a crosscheck we also derived a histogram MF at $z\sim4$ directly
from the masses of the $z\sim4$ sample using the search volume for the
$B$-dropouts. This straightforward process gives a MF that is
identified in Figure \ref{MFfig} as the ``direct MF'' for comparison
with the uncorrected ``bootstrap MF''.

To correct the MFs for incompleteness at
$\mathcal{M}<10^{8.5}\rm{M_\odot}$, we re-derive the MFs but now
including fainter ($-18<\rm{M_{UV,1500}}<-15$) sources.  We
extrapolate the observed log($\mathcal{M}$)--M$_{UV,1500}$ relation to
lower luminosities and assume that the low-luminosity scatter is
similar to the scatter around $M_{UV,1500}\sim-18.5$. The resulting
errors on the corrected points include an added uncertainty that is
typically about 30--40\%.
This accounts for the LFs uncertainties (specially the faint end
slope) and the large scatter about the relations. Other sources of
uncertainty may remain but further assessment is needed to fully
evaluate them.  Regardless, the current corrections must make these
corrected MFs a better estimate of the true MFs.  Applying the
completeness corrections is a crucial step and significantly changes
the slope of the MF at lower mass.  The corrected MFs are shown in
Figure \ref{MFfig} by the solid points and the solid color band, and
are referred to as the \emph{bootstrap} MFs.

\emph{Analytic MFs}: The best fit $\mathcal{M}$--M$_{UV,1500}$
relation in Figure \ref{mvsMUV} at $z\sim4$ is:
$\mathcal{M}\propto\rm{L}_{UV,1500}^{1.7(\pm0.2)}$.  We combine this
relation with the same LFs as above to derive the \emph{analytic}
MFs (Figure \ref{MFfig}). The large scatter needs to be accounted for
in order to generate a realistic estimate. The mass distribution at a
given luminosity is assumed to be log-normal with a standard deviation
of $0.5\,$dex, centered on the fitted
log($\mathcal{M}$)--M$_{UV,1500}$ relation. The resulting relation,
normalized by the LF, is integrated over L to get the \emph{analytic}
MFs.

At lower masses, the slope of the MF is set by the faint-end slope of
the UV-LFs and of the log($\mathcal{M}$)--M$_{UV,1500}$ relation.
Extrapolating this $\mathcal{M}$--M$_{UV,1500}$ relation to lower
luminosities results in steep low-mass slopes for the MFs of
$-1.43(\pm0.11)$, $-1.39(\pm0.11)$, $-1.44(\pm0.15)$, and
$-1.55(\pm0.21)$ at $z\sim4,~5,~6,$ and 7. These MF slopes are
slightly flatter than the UV LF slopes
($\alpha=-1.7--2$:~\citealt{bouw10d}). They are in good agreement with
our completeness-corrected \emph{bootstrap} MFs, and so provide a
useful ``sanity check'' on those results. The standard deviation of
0.5$\,$dex in the log($\mathcal{M}$)--M$_{UV,1500}$ relation results
in a slightly enhanced number density at the high-mass end compared to
a case with no scatter.  Our corrected MFs are considerably steeper
than other MF determinations at high redshift \citep{star09} that do
not apply completeness corrections.  Truncating the analytic
$\mathcal{M}$--M$_{UV,1500}$ relation at M$_{1500}<-18$ (to represent
survey incompleteness) results in analytic MFs that are in good
agreement with the non-corrected \emph{bootstrap} MFs at low masses.

\emph{Comparisons to simulated MFs}: The MFs derived here are
substantially steeper at low masses than what has been found in the
past at these redshifts. This is not unexpected given that it has not
been typical to correct for incompleteness.  While the corrections are
uncertain in magnitude the sign of the correction is
not. Interestingly, even with the corrected and steeper slopes, the
observed MFs are quite different from what is seen in recent
simulations (e.g. \citealt{choi10}; Finlator et al. in preparation).
The simulated MFs are steeper, with dramatically more low-mass sources
than we find.

\section{Stellar Mass Density at $z\sim4,~5,~6,~\rm{and}~7$}

\begin{figure*}
  \centering

  \includegraphics[width=1.\textwidth]{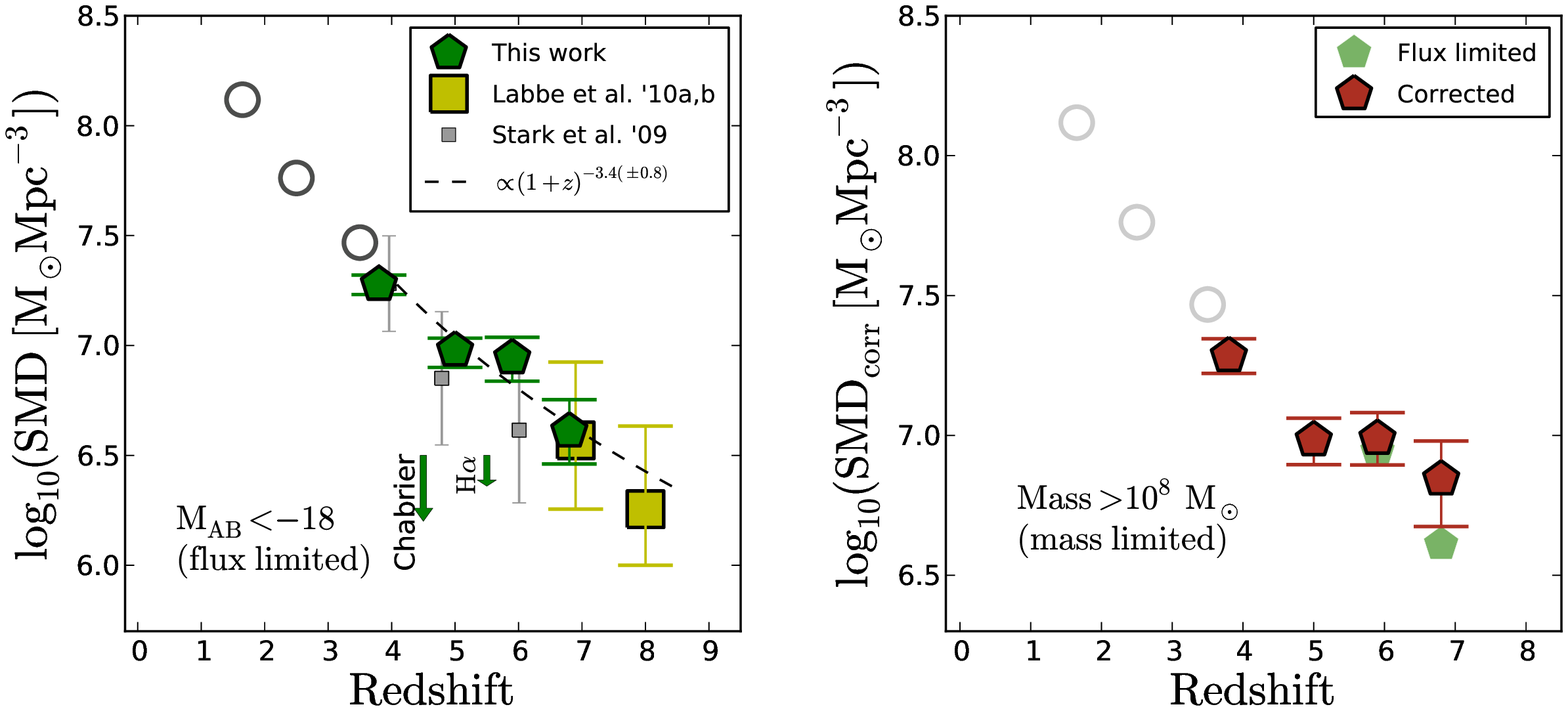}

  \caption{(\emph{left}) Stellar Mass Density as a function of
    redshift for sources brighter than
    M$_{\rm{UV,1500,AB}}=-18$. These SMD values are derived by
    integrating the uncorrected \emph{bootstrap} MFs in Figure
    \ref{MFfig} to the faint luminosity limit M$_{UV,1500}=-18$ at
    $z=4,~5,~6,~\rm{and}~7$.  For comparison, we show the SMD
    determinations from \cite{star09} corrected from their original
    M$_{UV,1500}=-20$ limit to our M$_{UV,1500}=-18$ limit (see
    text). The low-redshift open circles were derived by integrating
    the \cite{marc09} MFs between
    $\rm{8.3<log_{10}(\mathcal{M}/M_\odot)<13}$ and multiplying by 1.6
    to match the Salpeter IMF. A constant SFH and 0.2$\,$Z$_\odot$
    metallicity was assumed to derive the masses at $z\gtrsim4$. The
    effect of a possible 20\% correction due to contamination by
    H$\alpha$ is shown, as is the effect of using a different IMF. The
    integrated mass growth we derive with cosmic time is well fit by
    $\rm{log_{10}(SMD)\propto(1+z)^{-3.4\pm0.8}}$. (\emph{right}) As
    for the left panel but now to a fixed mass limit
    $>10^{8}~\rm{M}_\odot$.  The SMD to a fixed mass limit is compared
    to the flux limited values from the left panel.  The differences
    are relatively small (see text). However, the importance of being
    able to utilize the completeness-corrected MFs will increase as
    improved, deeper data becomes available and we can push to lower
    mass limits.}

  \label{SMDfig} 
\end{figure*} 

\begin{deluxetable*}{llccccc}
  \tablecolumns{7}
  \tablewidth{0pt}
  \tablecaption{Summary of Results. \label{SMDtable}}
  \tabletypesize{\scriptsize}
  \centering
  \tablehead{
  &&\colhead{$<z>=3.8$}&\colhead{$<z>=5.0$}&\colhead{$<z>=5.9$}
   &\colhead{$<z>=6.8$}&\colhead{$<z>=8.0$}\\
  }
  \startdata
  \multicolumn{7}{c}{COMPLETENESS CORRECTED MASS FUNCTIONS}\\ 
  \multicolumn{2}{l}{$\rm{log_{10}(\mathcal{M}/M_\odot)}$} & 
  \multicolumn{5}{c}{$\rm{log_{10}}(dN/dlog_{10}(\mathcal{M}/M_\odot)/Mpc^{3})$} \\
  \multicolumn{2}{l}{[7.5 - 8.0]} & $-1.90(^{+.11}_{-.12})$ & $-2.21(^{+.18}_{-.20})$ & $-2.09(^{+.23}_{-.24})$ &  $-2.15(^{+.41}_{-.39})$ & \nodata\\
  \multicolumn{2}{l}{[8.0 - 8.5]}   & $-1.96(^{+.12}_{-.12})$ & $-2.27(^{+.18}_{-.20})$ & $-2.16(^{+.23}_{-.23})$ &  $-2.23(^{+.37}_{-.38})$ & \nodata\\ 
  \multicolumn{2}{l}{[8.5 - 9.0]}   & $-2.30(^{+.10}_{-.11})$ & $-2.60(^{+.15}_{-.19})$ & $-2.55(^{+.20}_{-.21})$ &  $-2.70(^{+.34}_{-.36})$ & \nodata\\
  \multicolumn{2}{l}{[9.0 - 9.5]}   & $-2.53(^{+.10}_{-.11})$ & $-2.84(^{+.17}_{-.17})$ & $-2.81(^{+.19}_{-.22})$ &  $-3.01(^{+.34}_{-.34})$ & \nodata\\
  \multicolumn{2}{l}{[9.5 - 10.0]}  & $-3.14(^{+.12}_{-.13})$ & $-3.46(^{+.17}_{-.21})$ & $-3.52(^{+.21}_{-.23})$ &  $-3.80(^{+.36}_{-.37})$ & \nodata\\
  \multicolumn{2}{l}{[10.0 - 10.5]} & $-3.80(^{+.20}_{-.23})$ & $-4.12(^{+.22}_{-.27})$ & $-4.22(^{+.28}_{-.31})$ &  $-4.53(^{+.39}_{-.61})$ & \nodata\\
  \multicolumn{2}{l}{[10.5 - 11.0]} & $-4.43(^{+.26}_{-.46})$ & $-4.81(^{+.33}_{-.45})$ & $-4.97(^{+.38}_{-.70})$ & \nodata & \nodata\\

  \hline\hline\\

  \multicolumn{2}{l}{SMD ($\mathcal{M}>10^8\rm{M_\odot}$)} & 
  $19.27(_{-2.62}^{+2.88})$ &
  $9.64(_{-1.78}^{+1.88})$ & 
  $9.76(_{-1.91}^{+2.30})$ &
  $6.98(_{-2.26}^{+2.57})$ & \nodata\\
  $[\rm{10^6~M}_\odot]$&&&&&\\ 
  \\
  \tableline\\

 \multicolumn{2}{l}{SMD (M$_{1500}<-18$)} & $18.96(^{+1.94}_{-1.90})$ &
  $9.52(^{+1.27}_{-1.58})$ & $8.79(^{+2.11}_{-1.91})$ & 
  $4.08(^{+1.59}_{-1.19})$ & $1.8(^{+0.7}_{-1.0})$\\

  $[\rm{10^6~M}_\odot]$&&&&&\\
  Best Fit & \multicolumn{6}{c}{
    $\rm{log_{10}(SMD(z)/[M_\odot~Mpc^{-3}])}=7.00(^{+0.04}_{-0.05})-3.35(^{+0.82}_{-0.94})\times\rm{log_{10}}(\frac{1+z}{6})$}\\

\enddata
\end{deluxetable*}

The MFs can be integrated to determine the SMD of the universe at high
redshift.  First, we integrate the (uncorrected) bootstrap MFs to
determine the SMD at $z=4,~5,~6,~\rm{and}~7$ to faint-luminosity
limits (M$_{1500}<-18$ -- Table \ref{SMDtable}; Figure \ref{SMDfig}
left).  Fitting the SEDs using the observed fluxes rather than upper
limits allows us to reach lower limits than those of \cite{star09} at
$z=4-6$.  To compare to those results, we correct their original
M$_{1500}=-20$ limit to our M$_{1500}=-18$ limit by adding
$0.18,~0.22,$ and 0.32$\,$dex at $z\sim4,~5$, and 6 respectively. We
derive an integrated mass growth across cosmic time that is well fit
by the function $\rm{log_{10}(SMD)\propto(1+z)^{-3.4\pm0.8}}$.  The
effect of a different IMF and of a potential contamination by 20\%
$H\alpha$ at $z\sim 5-6$ is also shown on Figure \ref{SMDfig}.

A major result of this paper is the derivation of MFs corrected for
incompleteness at low masses. To utilize the (more representative and
accurate) corrected MFs in deriving the SMD of the universe at high
redshift we need to integrate to a fixed mass limit.  We choose
$\mathcal{M}>10^{8}~\rm{M}_\odot$ to extend to the limit of our
corrected data.  The right panel of Figure \ref{SMDfig} shows these
results, and compares them with the estimates to a fixed luminosity
limit (see also Table \ref{SMDtable}).

The differences are relatively small, less than we expected given the
size of the corrections.  A number of effects impact the comparison.
For the flux-limited ($M_{AB}<-18$) sample, we include relatively
bright sources that have masses $>10^8~\rm{M}_\odot$. In the
mass-limited case, we remove those galaxies, and instead include faint
sources with large masses.  Depending on the limits adopted and the
shape of the LF, this effect is more or less important. At $z\sim7$,
the fainter L$^*$ and the steep faint end slope make the difference
between what is removed and what is added larger. For the brighter
L$^*$ at lower redshifts, however, the difference is almost null.

Nonetheless, the derivation of mass-limited SMDs that appropriately
include corrections at low masses is preferred and will become
increasingly important as we push our measurements of the MF to lower
limits.

\section{Key Results}

We derive stellar masses from SED fits to HST+Spitzer data for over
400 $z\sim4-7$ galaxies. We determine the $\mathcal{M}$--L$_{UV}$
relation and find it to be steep
(log$(\mathcal{M})\propto1.7(\pm0.2)\rm{log(L}_{\mathit{UV}})$) with
large intrinsic scatter; the sample variance is $\sim0.5\,$dex at the
bright end. We derive mass functions by combining the $\mathcal{M}/$L
results with published deep UV luminosity functions at $z\sim4-7,$ and
correct them for incompleteness.  The corrected mass functions are
steeper ($\alpha\sim-1.4$ to $-1.6$) than found previously, but still
far less steep than those from recent hydrodynamical simulations. The
integrated stellar mass density of the universe is derived at
$z\sim4,~5,~6,$ and 7 to $\mathcal{M}\sim10^8M_\odot$.
	
\acknowledgments We acknowledge insightful discussions with Kristian
Finlator, Casey Papovich, Daniel Schaerer, and Daniel Stark. We thank
Ken Nagamine, Junhwnan Choi and Kristian Finlator for access to their
simulations.  We acknowledge support from HST-GO10937, HST-GO11563,
HST-GO11144, and a Fulbright-CONICYT scholarship (V.G.).  I.L. is
supported by NASA through a Hubble Fellowship grant \#51232.01-A
awarded by the Space Telescope Science Institute.

\bibliographystyle{apj}

\end{document}